\documentclass[twocolumn,english,prl]{revtex4}
\usepackage[T1]{fontenc}
\usepackage[utf8]{inputenc}
\usepackage{color}
\usepackage{amsmath}
\usepackage{amssymb}
\usepackage{graphicx}
\usepackage{esint}

\usepackage[caption=false]{subfig}
\usepackage[english]{babel}
\usepackage{amsfonts}

\newcommand{\pb}{\textbf{p}}

\newcommand{\Ab}{\textbf{A}}
\newcommand{\Bb}{\textbf{B}}
\newcommand{\Eb}{\textbf{E}}

\begin{document}
    \title{Fixing E-field divergence in strongly nonlinear wakefields in homogeneous plasma}
	\author{Lars Reichwein}
	\email{lars.reichwein@hhu.de}
	\author{Johannes Thomas}
	\affiliation{Institut f\"{u}r Theoretische Physik I, Heinrich-Heine-Universit\"{a}t D\"{u}sseldorf D-40225 D\"{u}sseldorf, Germany}
	
	\author{Anton Golovanov}
	\author{Igor Yu. Kostyukov}
	\affiliation{Lobachevsky State University of Nizhny Novgorod, 603950 Nizhny Novgorod, Russia}
	\affiliation{Institute of Applied Physics RAS, 603950 Nizhny Novgorod, Russia}

	\author{Alexander Pukhov}
	\affiliation{Institut f\"{u}r Theoretische Physik I, Heinrich-Heine-Universit\"{a}t D\"{u}sseldorf D-40225 D\"{u}sseldorf, Germany}
	
	\date{\today}
	\begin{abstract}
		Available analytical wakefield models for the bubble and the blow-out regime of electron-plasma acceleration perfectly describe important features like shape, fields, trapping ratio, achievable energy, energy distribution and radial emittance. As we show, for wakefields with an extremely small amplitude these models fail to describe the accelerating electric field and its divergence in the wakefield rear. Since prominent parameter regimes like the Trojan horse regime of photocathode injection exhibit this feature, it is of great importance to work out analytical models that fix this problem; one possible model is introduced in this work. Using a phenomenological theory, we are able to better describe the divergence of the electric field and the bubble shape.
	\end{abstract}
	\maketitle

	
	\section{Introduction}
	Plasma is an ionized medium that supports electric fields that are several orders of magnitude higher than those in traditional solid-state accelerating structures (metal, warm or superconducting). The accelerating field in plasmas can be excited either by an intense laser pulse with wavelength $\lambda_L$, duration $\tau$ and focal spot size $R$ \cite{Tajima1979}, or by a charged particle beam with length $\sigma_z$, radius $\sigma_r$ and density $n_b$ \cite{Chen1985, Rosenzweig1988}. 
	
	In plasmas with homogeneous density $n_p$, the wakefield breaks as soon as the laser pulse intensity reaches a certain threshold value and the normalized laser amplitude $a_0>1$. If $a_0 > 4$, $\lambda_p > R > 2\lambda_L$ and if the laser pulse perfectly fits into the first half of the plasma period, a solitary electronic cavity, called the bubble, is formed \cite{Pukhov2002, Jansen2014, Pukhov2006, Lu2006, Lu2007}. It is a nearly spherical region with uniform accelerating fields that propagates with almost the speed of light $c$ \cite{Kostyukov2004}. In homogeneous plasma, it traps background electrons at its tail and accelerates them to high energies. The major features that characterize the bubble regime are the quasi-monoenergetic spectrum of the accelerated electrons and the quasi-static laser pulse which propagates many Rayleigh lengths in homogeneous plasma without significant diffraction.
	
	If the wakefield is excited by a thin $\sigma_z\approx\sqrt{2}k_p^{-1}\gg \sigma_r$ and dense $n_b>n_p$ charged particle beam, a spherical ion column, the so-called blow-out, is created. It is a structure similar to the laser driven bubble with a comparable accelerating field $E^+$ in its rear part. 
	An important parameter for the energy gain of trapped electrons is the transformer ratio $T=E^+/E^-$. It describes the maximum energy gain of the witness bunch at simultaneous energy loss of the driver beam in its peak decelerating field $E^-$ \cite{Lu2006, Hogan2005, Huang2017}.
	If the total charge of the witness beam exceeds a certain threshold, the plasma cavity structure both in the bubble and blow-out regime is reshaped and the effective accelerating field is modified. This in turn affects final beam properties like maximum energy, energy spread, total charge, transverse emittance, the spatial extension in length and width, but also general acceleration parameters like the transformer ratio \cite{Golovanov2016, Lu2007, Gordienko2005, Tzoufras2008, Tzoufras2009, Vafaei-Najafabadi2014, Couperus2017}. To optimize all characteristic witness-beam parameters, an effective loading technique is necessary.

	A promising method to control beam loading is the ionization injection technique. This method produces witness electron beams with sub-fs temporal duration, a very high peak current (several kA), energy spreads well below 1\% for beams with energies in the multi-GeV range and an excellent transverse emittance (tens of nm\,rad) \cite{Huang2017, Baxevanis2017, Wang2018, Gonsalves2017, Martinez2017, Tooley2017}. The ionization injection requires a small amount of higher-Z gas, added to the gas used for acceleration \cite{Pak2010, Tochitsky2016}. The ionization process starts as soon as the local field strength exceeds the ionization threshold. Therefore it can be triggered by the laser pulse exciting the wakefield \cite{Wang2018, Lee2017, Joshi2017}, by the wakefield itself \cite{Martinez2013}, by transversely colliding laser pulses \cite{Li2013, Chen2014}, or by a second, trailing, laser pulse. In the latter case, wakefield excitation and ionization are independent processes, which allows a precise manipulation of the phase space distribution of trapped electrons and thus a generation of ultra-low emittance electron beams \cite{Chen2012, Bourgeois2013, Shaw2014, Albert2016a, Schroeder2015, Schroeder2018, Yu2014, Yu2016}. If the driving laser pulse is replaced by a short electron beam, the Trojan horse regime (THWFA) of underdense photocathode plasma wakefield acceleration (PWFA) is reached \cite{Hidding2012, Hidding2012a}. It is best suited for decoupling the electron bunch generation process from the excitation of the accelerating plasma cavity. Since the blow-out structure is largely immune to shot-to-shot driver bunch characteristics variations, the acceleration process can be controlled via plasma density modulations, while the charge of the witness bunch can be tuned by the release laser intensity. The combination of the non-relativistic intensities required for tunnel ionization ($10^{14}$ W/cm$^2$), a localized release volume as small as the laser focus, the greatly minimized transverse momenta, and the rapid acceleration leads to dense phase space packets. In homogeneous plasma they can have ultra-low normalized transverse emittance of the order of nm\,rad and a minimal energy spread in the 0.1\% range. Achieved electron energies after acceleration in the bubble or blow-out regime are in the range of 1 GeV to 8 GeV \cite{Huang2017, Leemans2006, Blumenfeld2007, Kim2013, Wang2013, Leemans2014, Vafaei-Najafabadi2016, Albert2016, Gonsalves2019}, while simulations and planned experiments aim for the generation of 10 GeV beams in a single acceleration stage \cite{Davidson2017, Gonsalves2017, Katsouleas2006a, Daniels2017, Joshi2018}.

	In the following section, we first recap fields and forces in the bubble for various plasma configurations and bubble shapes---including small amplitude wakefields similar to those known from Trojan horse regime. Afterwards, in section \ref{sec:model}, we  present an analytical model for small amplitude wakefields, beginning with an approximation of the electron layer. Already available theoretical bubble models for THWFA lose their applicability towards the bubble back, since there the electric field exhibits a divergence.
	We derive field approximations for homogeneous plasma, introducing an artificial trick to model the field divergence at the bubble back phenomenologically.
	\begin{figure*}
		\subfloat[]{\label{TroHo_0_Ex_qsm}\includegraphics[width=0.3\textwidth]{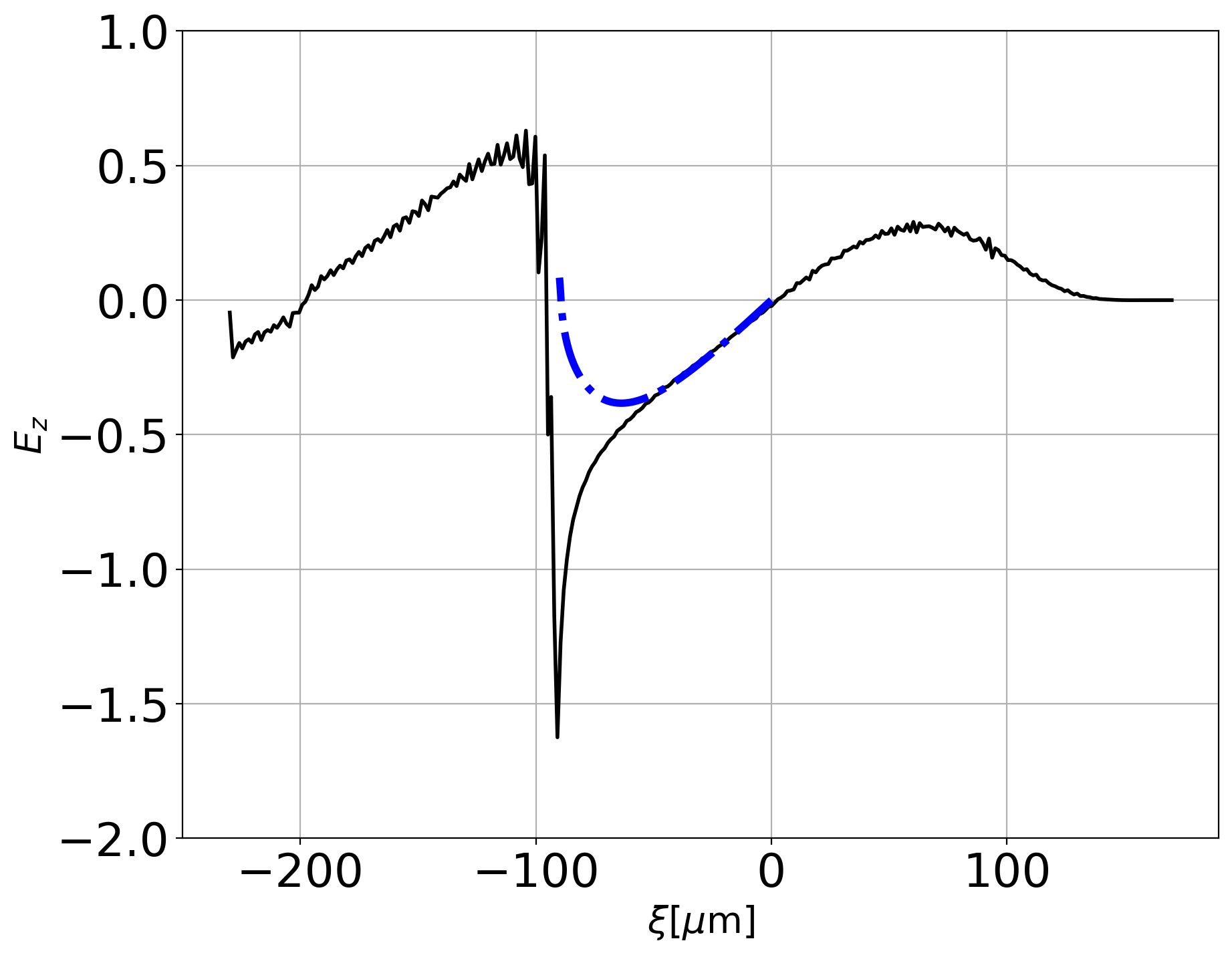}}
		\subfloat[]{\label{TroHo_3_Ex_qsm}\includegraphics[width=0.3\textwidth]{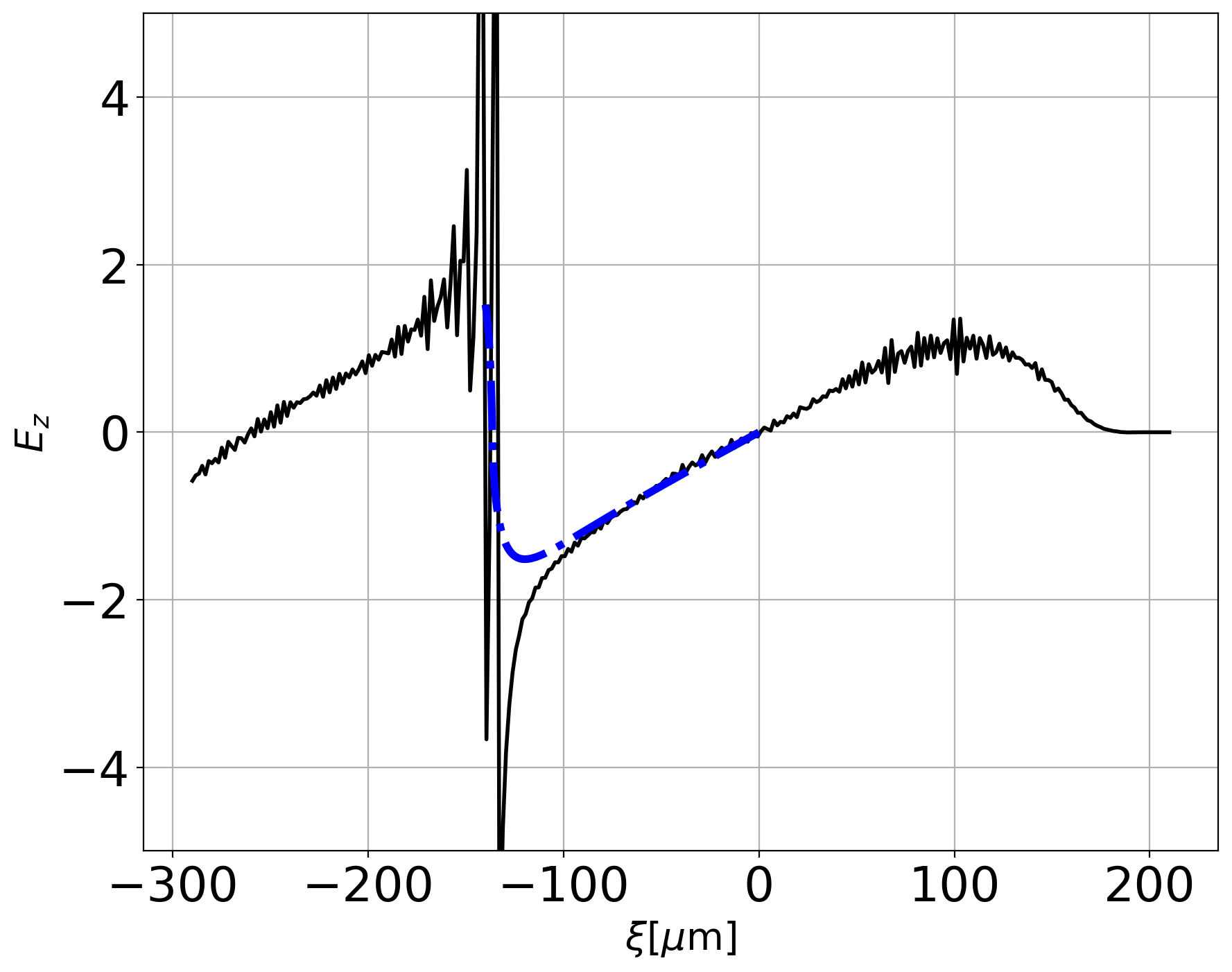}}
		\subfloat[]{\label{TroHo_4_Ex_qsm}\includegraphics[width=0.3\textwidth]{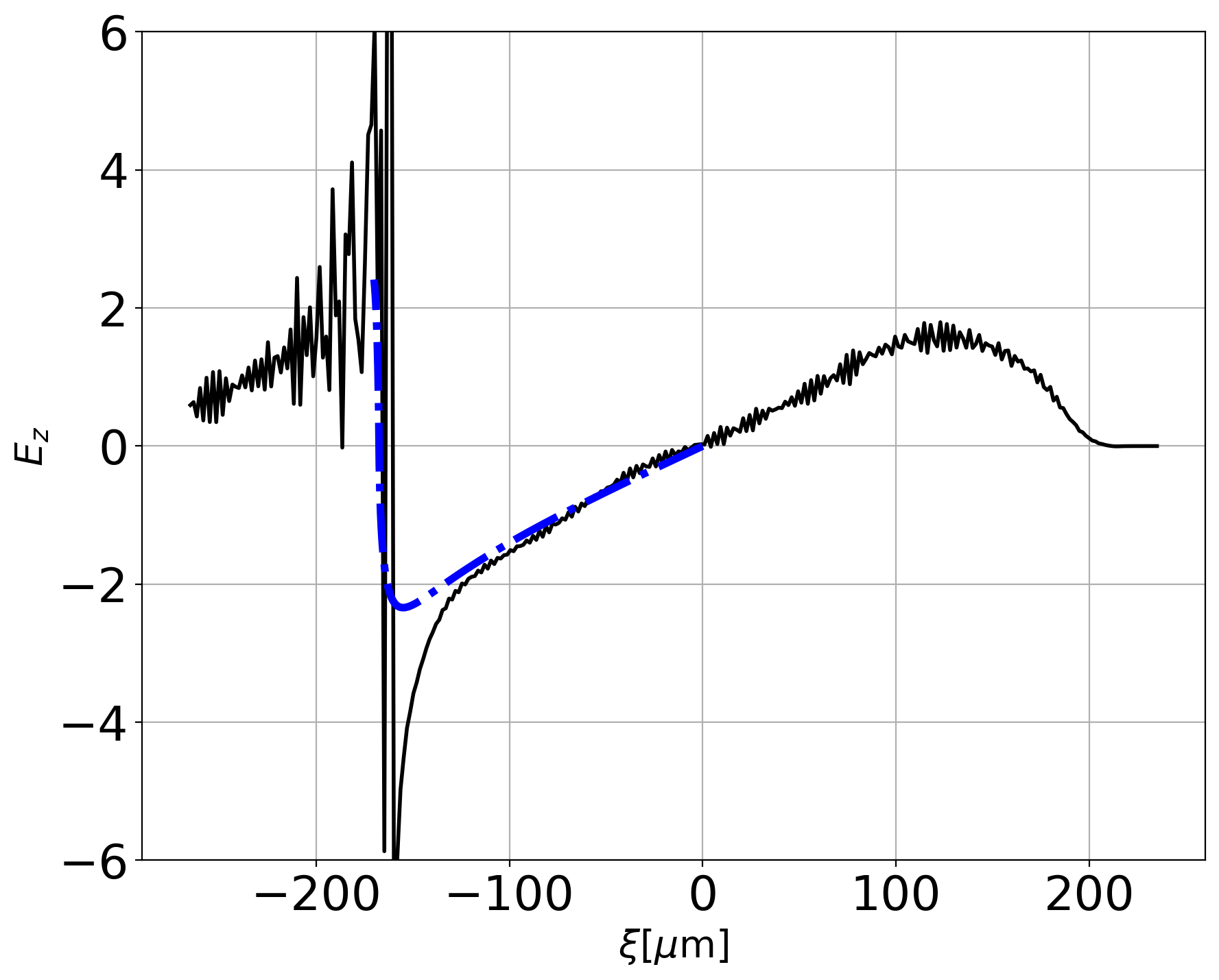}}
		\caption{\label{fig:fieldsBroken}$E_z$-field in dependence of the longitudinal coordinate $\xi$ from PIC simulations introduced in Fig.~\ref{fig:PICfit} (black lines). The analytical model without repair function from Eq.~(\ref{Ez_qsm}) (blue dashed lines) describes the fields well in the bubble mid, but not towards its back, in the region around $\xi \leq -100$ $\mu$m.}
	\end{figure*}

	
	\section{Theoretical model: Forces}\label{sec:forces}
	In the following we derive the radial and longitudinal force components acting on a test electron moving inside the bubble for various cases regarding plasma density and bubble shape.
	Throughout the derivation we  use normalized units, where time is normalized to the inverse plasma frequency $\omega_p^{-1} = \sqrt{4\pi e^2 n_e / m_e}^{-1}$, lengths to $k_p^{-1} = c / \omega_p$, kinetic momenta to $m_e c$, energies to $m_e c^2$, fields to $m_e c \omega_p / e$, charges to the elementary charge $e$, masses to the electron's rest mass $m_e$ and potentials to $m_e c^2 /e$. Throughout the simulations, we use the plasma wavelength of $\lambda_p = 200$ $\mu$m.

	Starting point for all of the following calculations is the quasi-static wakefield potential \cite{Kostyukov2004}, which is generally given as
	\begin{align}
		\Psi = A_z - \varphi \; .
	\end{align}
	As the bubble is moving with velocity $V_0$ close to the speed of light, we express all functions in dependence of the coordinate $\xi = z - t$. In this co-moving frame with cylindrical symmetry and distance to the symmetry axis $r$, the magnetic and electric field are
	\begin{align}
		\Bb &= \nabla\times\Ab = \left(\frac{\partial A_r}{\partial\xi} -\frac{\partial A_z}{\partial r} \right)\hat{e}_\varphi \; , \\
		\Eb &= \nabla\Psi -\hat{e}_z \times \Bb\; .
	\end{align}
	The written-out electric field components are connected to the wakefield potential via
	\begin{align}
		& E_z = \frac{\partial \Psi}{\partial \xi} \; , && E_r = \frac{\partial \Psi}{\partial r} + B_\varphi \; .
	\end{align}
	In general, the corresponding force components become
	
	\begin{align}
	F_z &= -E_z -\frac{p_r}{\gamma}B_\varphi \; ,\\
	F_r &= -\frac{\partial \Psi}{\partial r} -\left(1 -\frac{p_z}{\gamma}\right)B_\varphi \; ,
	\end{align}
	where $\gamma = \sqrt{1+|\pb|^2}$ and $\pb$ is the electron's kinetic momentum.

	For the sake of generality, we consider an arbitrary radial plasma profile $\rho(r)$ and follow the models of \cite{Thomas2016, Golovanov2017b} for the definition of the wakefield potential in the bubble or blow-out regime in dependence on the coordinates $\xi$ and $r$:
	
	\begin{align}
		\Psi(\xi,r) = \int_0^r \frac{S_I(r')}{r'} \; dr' +\Psi_0(\xi) \; .
	\end{align}
	Here,
	\begin{align}
		S_I(x) = \int_0^{x} \rho(r)r \; dr > 0 \label{eq:intSrc}
	\end{align}
	is the integral source, and
	
	\begin{align}
		\Psi_0(\xi) = -\int_0^{r_b}\frac{S_I(r)}{r} \; dr -\frac{S_{I,b}\beta(r_b)}{2} \; .
	\end{align}
	From now on, we use $S_{I,b}$ as a shorthand notation for $S_I(r_b)$.
	The parameter $\beta$ is in turn defined as
	
	\begin{align}
		\beta(r_b) = 2\int_0^\infty\frac{\epsilon dx}{1+\epsilon x}\frac{F_0(x) +\epsilon F_1(x)}{1+\epsilon F_1(0)} \; ,
	\end{align}
	where $r_b(\xi)$ is the bubble radius, $\epsilon=\frac{\Delta}{r_b}$ is the relative sheath width and
	\begin{align}
		F_n(x) = \int_x^\infty y^n g(y)\; dy
	\end{align}
	is the generalized moment of the function $g(y)$ which describes the the shape of the electron sheath at the bubble's boundary, and $F_0(0) = 1$ is assumed.
	\subsection{Arbitrary plasma}
	For a plasma with arbitrary radial density profile $\rho(r)$, we have the integral source (\ref{eq:intSrc}) and the field components are
	
	\begin{align}
		E_z &= -\left(\frac{S_{I,b}}{r_b} +\frac{S_{I,b}\beta'(r_b)}{2} +\frac{\rho(r_b)\beta(r_b)r_b}{2}\right)r_b' \; , \label{Ez_qsm} \\
		B_\varphi(\xi,r) &= -\frac{r}{2}\frac{\partial E_z}{\partial\xi} -\Lambda(r,\xi) \; , \\
		E_r &= \frac{\partial \Psi}{\partial r} +B_\varphi = \frac{S_I(r)}{r} + B_\varphi \; .
	\end{align}
	With the integral current density
	\begin{align}
	    \Lambda(r,\xi) = -\frac{1}{r}\int_0^r J_z(\xi,r')r'dr' \;\label{dari} 
	\end{align}
	this leads to the following force components in the transverse and longitudinal directions:
	\begin{align}
	   F_z &= -E_z -\frac{p_r}{\gamma}B_\varphi \; , \\
	   F_r &= -\frac{S_I(r)}{r} -\left(1 -\frac{p_z}{\gamma}\right)B_\varphi \; .
	\end{align}
	
	\subsection{Homogeneous plasma}\label{homogeneous}
	In the case of a homogeneous ($\rho(r) = 1$, $S_I(r) = r^2 / 2$) plasma, the components of the electric field exhibit the form
	
	\begin{align}
		E_r &= \frac{r}{2} +B_\varphi \; , \\
		E_z &= -\frac{r_br_b'}{2}\left(1 +\beta(r_b) +\frac{r_b\beta'(r_b)}{2}\right) \; ,
	\end{align}
	leading to 
	\begin{align}
		F_z &= \frac{r_br_b'}{2}\left(1 +\beta(r_b) +\frac{r_b\beta'(r_b)}{2}\right) +\frac{p_r}{\gamma}B_\varphi \; , \\
		F_r &= -\frac{r}{2} -\left(1 -\frac{p_z}{\gamma}\right)B_\varphi \; .
	\end{align}
	
	\subsection{Homogeneous plasma and spherical bubble with thin sheath}
	If the bubble, in addition, exhibits a spherical shape, we still have $\rho(r) = 1$ and $S_I(r) = r^2 / 2$, but now set $r_b = \sqrt{R^2 - \xi^2}$ with $r_b' = - \frac{\xi}{r_b}$. If we further assume that $\beta = \beta' = 0$, the electric and magnetic field components reduce to the well known form
	\begin{align}
		& E_z = \frac{\xi}{2} \; , && B_\varphi = -\frac{r}{4} - \Lambda(r, \xi) \; , && E_r = \frac{r}{4} - \Lambda(r, \xi) \; .
	\end{align}
	In this case, the accelerating and focusing forces are
	\begin{align}
		F_z &= -\frac{\xi}{2} +\frac{rp_r}{4\gamma} + \frac{p_r}{\gamma} \Lambda(r,\xi) \; , \\
		F_r &= -\frac{r}{4}\left(1 +\frac{p_z}{\gamma}\right) - \frac{1}{2\gamma^2} \Lambda(r,\xi) \; .
	\end{align}
	Since the particles undergo fast betatron oscillations, we can assume that $\langle r p_r \rangle = 0$ and since $p_z \approx \gamma$ after sufficient acceleration, the terms above can be simplified to
	\begin{align}
		 & F_z \approx - \frac{\xi}{2}\; , && F_r \approx - \frac{r}{2} \;
	\end{align}
	for trapped electrons. These forces are already known from simple models (compare e.g. \cite{Kostyukov2004, Kostyukov2010, Thomas2014}). They do, however, require a lot of restrictions regarding the current distribution in the electron sheath, the bubble shape, the bubble symmetry, and the background plasma density profile.

	\subsection{Homogeneous plasma and stretched bubble with thin sheath}
	
	Leaving some of the restrictions mentioned above behind, we lastly want to consider the more general case of a stretched bubble with small amplitude $R\ll\lambda_p$ as it can be found in \cite{Hidding2012a}. If we generalize the bubble radius $r_b = R\sqrt{1 - (\xi/b)^2}$, $b>R$ to describe a stretched bubble shape, the electric field component in $\xi$-direction becomes $E_z = a \xi$ with $a \in \left(0, \frac{1}{2}\right)$. If we stay in the thin-sheath approximation $\beta = \beta' = 0$, we get
	\begin{align}
		& B_\varphi = -\frac{r}{2}a -  \Lambda(r,\xi) \; , && E_r = \frac{r}{2}(1-a) -  \Lambda(r,\xi) \; . 
	\end{align}
	The resulting force components are
	\begin{align}
		 F_z &= -a\xi +\frac{rp_r}{2\gamma}a + \frac{p_r}{\gamma}  \Lambda(r,\xi) \; , \\
		 F_r &= -\frac{r}{2}\left(1 -a +a\frac{p_z}{\gamma}\right) - \frac{1}{2\gamma^2} \Lambda(r,\xi) \; .
	\end{align}
	Similarly to the previous case, we still can assume that $\langle r p_r \rangle = 0$ and $p_z \approx \gamma$, thus
	\begin{align}
		& F_z \approx -a\xi \; , && F_r \approx -\frac{r}{2} \; .
	\end{align}
	This shows that the elongation of the wakefield does not change the focusing force, as long as the electrons are already trapped and fulfill the conditions of performing betatron oscillation and having high energy. The accelerating force, however, is dependent on the bubble elongation, as can be seen by the factor $a$ in the equation above.

    In the following section we derive the connection between the bubble elongation $b$ and the field modulation factor $a$.

	
	\section{Theoretical model: Electron sheath}\label{sec:model}
	In this section we first find an analytical approximation to the ODE that describes the electron sheath approximation function $r_b$. Afterwards, we  take a look into field approximations for homogeneous plasma.
	\subsection{General electron layer approximation}
	In general, we solve the ODE \cite{Thomas2016} 
	\begin{align}
	& A(r_b)r_b'' + B(r_b)\left(r_b'\right)^2 + C(r_b) = \frac{\Lambda(r_b(\xi),\xi)}{r_b} \label{ODE}
	\end{align}
	to find the electron sheath approximation function $r_b$. Here, the coefficient functions are
	\begin{align}
	A(r_b) &=  1 +\frac{S_{I,b}}{2} \notag \\ & + \left(\frac{\rho(r_b)r_b^2}{4} +\frac{S_{I,b}}{2}\right)\beta +\frac{S_{I,b}r_b}{4} \beta' \; , \label{A} \\
	B(r_b) &= \frac{\rho(r_b)r_b}{2} +\left[3\rho(r_b)r_b \notag +\rho'(r_b)r_b^2\right]\frac{\beta}{4} \\ &+[S_{I,b} +\rho(r_b)r_b^2]\frac{\beta'}{2} +S_{I,b}r_b\frac{\beta''}{4} \; , \label{B} \\
	C(r_b) = & \frac{S_{I,b}}{2r_b} \left[1 + \left(1 + \frac{S_{I,b}\beta}{2}\right)^{-2}\right] \; . \label{C}
	\end{align}
	\begin{figure}
		\subfloat[]{\label{TroHo_0_Ex_fit}\includegraphics[width=\columnwidth]{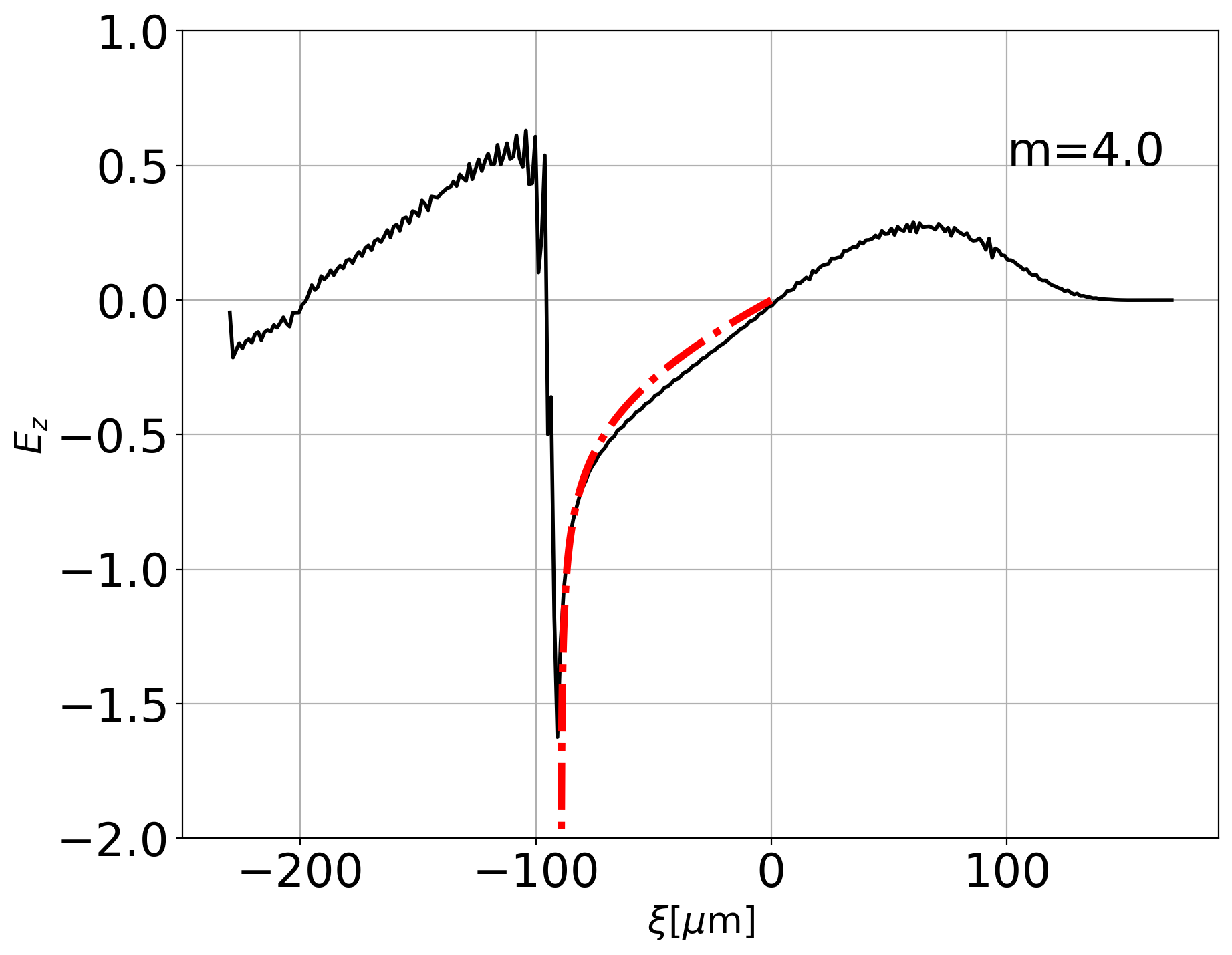}}\\
		\subfloat[]{\label{TroHo_3_Ex_fit}\includegraphics[width=\columnwidth]{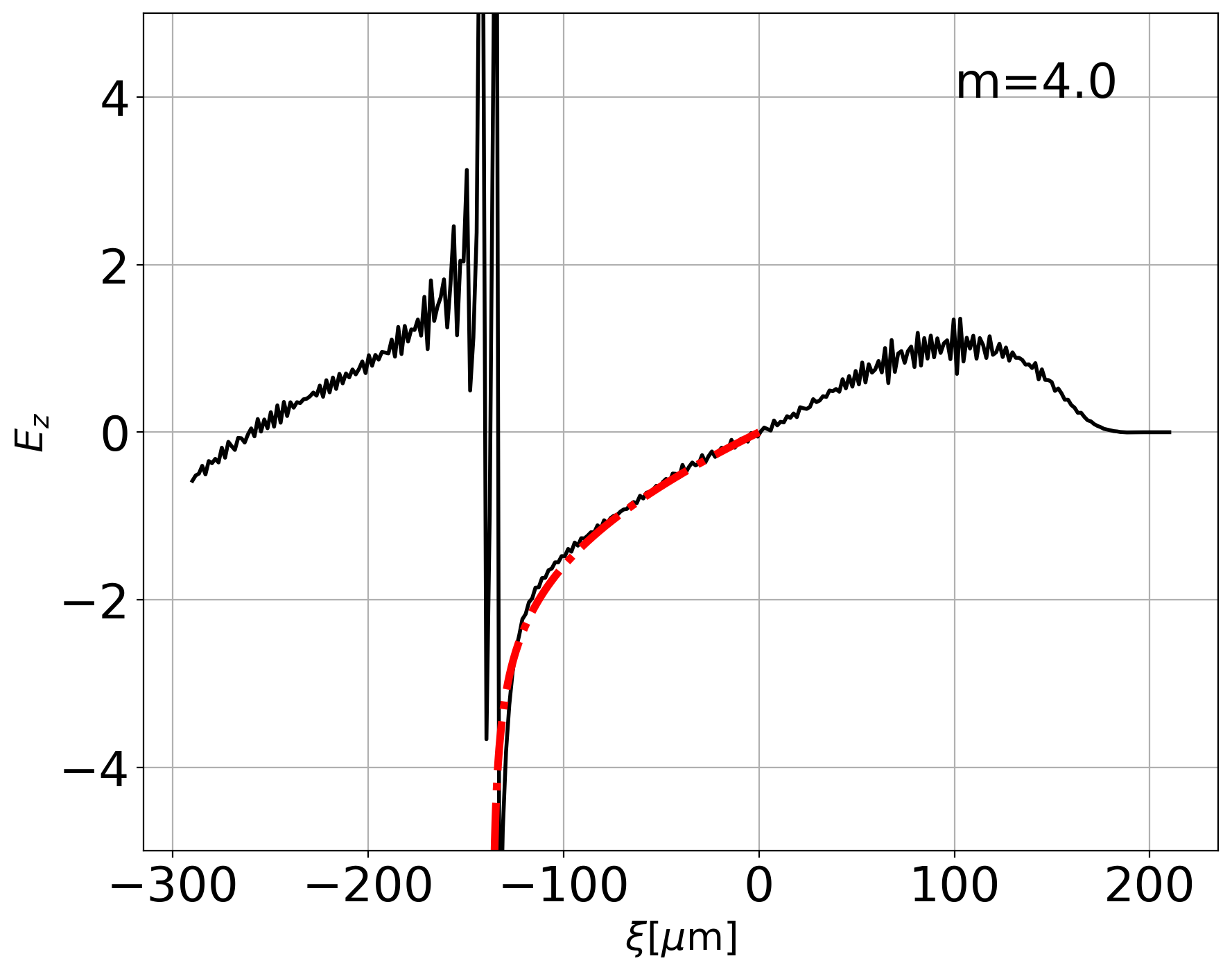}}\\
		\subfloat[]{\label{TroHo_4_Ex_fit}\includegraphics[width=\columnwidth]{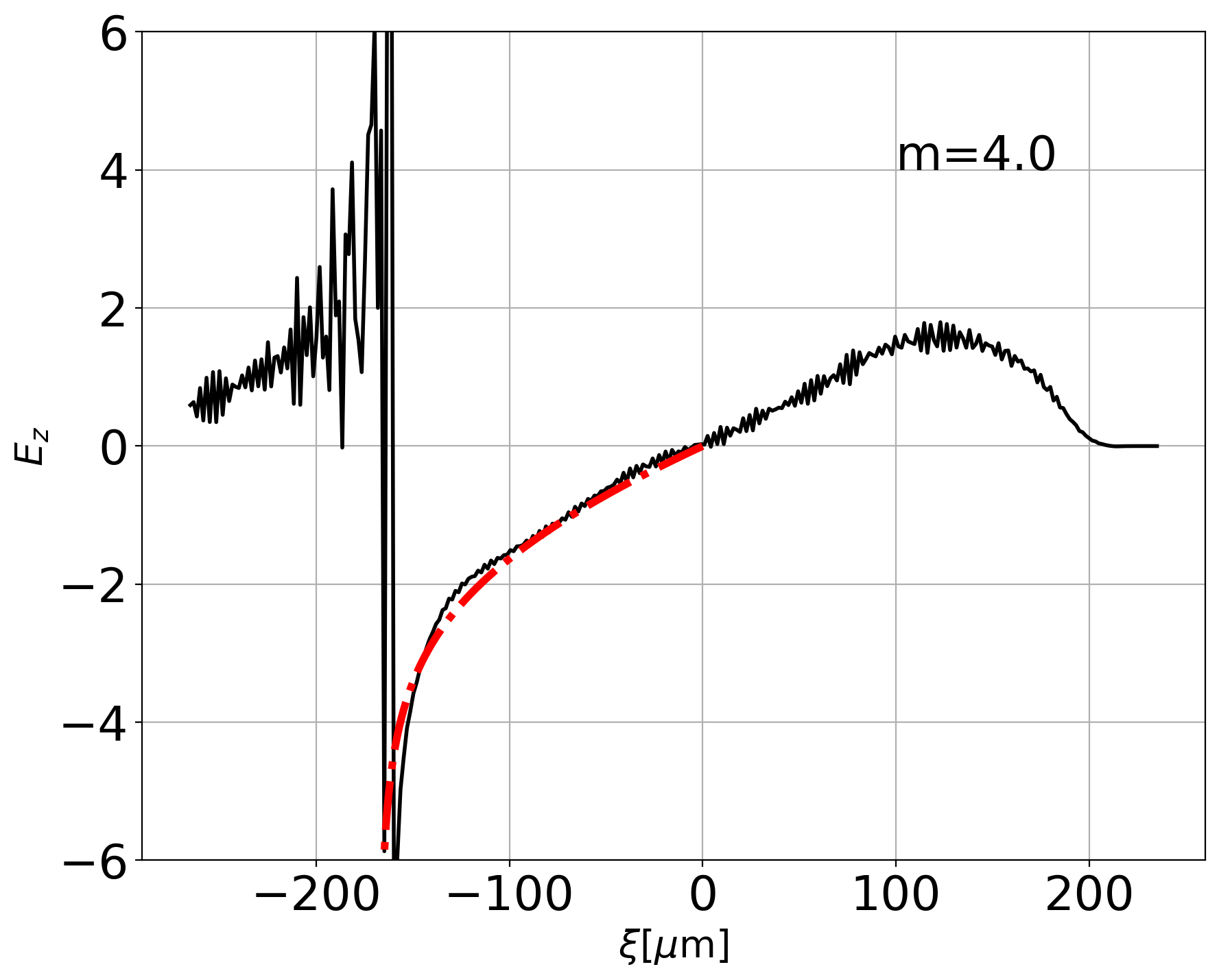}}
		\caption{\label{fig:fieldsRepaired}$E_z$-field of the bubble in dependence of the longitudinal coordinate $\xi$ (black lines). The analytical model with repair function (red dashed lines) is able to describe the fields well in the bubble mid and even towards its back, in the region around $\xi \leq -100$ $\mu$m.}
	\end{figure}

	If $\Delta\ll r_b$ and $\Delta\gg2r_b/S_{I,b}$, these functions can be simplified dramatically, so that 
	\begin{align}
	& A(r_b) \approx \frac{S_{I,b}}{2} \; , && B(r_b) \approx \frac{\rho r_b}{2} \; ,\\ & C(r_b) \approx \frac{S_{I,b}}{2 r_b} \; , && \Lambda(\xi) = \Lambda(r_b(\xi), \xi) 
	\end{align}
	for usual bubbles with rather large radii in the range of one plasma wavelength ($\lambda_p=2\pi k_p^{-1}$). Unfortunately, this approximation is not valid for cases similar to the Trojan Horse regime \cite{Hidding2012}, where the radius is in the order of $0.25\lambda_p\approx1.5 k_p^{-1}$ and thus $S_{I,b}/2$ is in the order of unity. To compensate this circumstance, we modify the coefficient functions phenomenologically to 
	\begin{align}
	& A(r_b) \approx 1 +\frac{S_{I,b}}{2} \; , && B(r_b) \approx \frac{\rho r_b}{2} \; , \\
	& C(r_b) \approx \frac{S_{I,b}}{2 r_b} \; , &&\Lambda(\xi) = \Lambda(r_b(\xi), \xi) \; . \label{reduced}
	\end{align}
	An analytical approximation to the ODE solution can be found for non-loaded ($\Lambda = 0$) bubbles and small $|\xi|$ near the bubble center (also compare \cite{Golovanov2016}). Here, 
	\begin{align}
	  & r_b \approx R \; , && r_b'\approx 0 \; , && r_b'' \approx - \frac{C}{A} \approx -\frac{S_{I,b}}{(2 +S_{I,b})R} \; ,
	\end{align}
	and
	\begin{align}
		S_{I,b} =  \frac{R^{2}}{2}
	\end{align}
	for homogeneous plasma, so that 
	\begin{align}
	& r_b \approx R -\frac{1}{1+ 4R^{-2}}\frac{\xi^2}{2R} \; . \label{ANA}
	\end{align}
	
	
	Comparing the solutions to Eqs. (\ref{A})-(\ref{C}) with the PIC results would show that they coincide with the simulations. The analytical expression would be sufficient near the bubble middle but could not describe the bubble back.
	
	\begin{figure}
		\subfloat[]{\label{TroHo_0_layer}\includegraphics[width=\columnwidth]{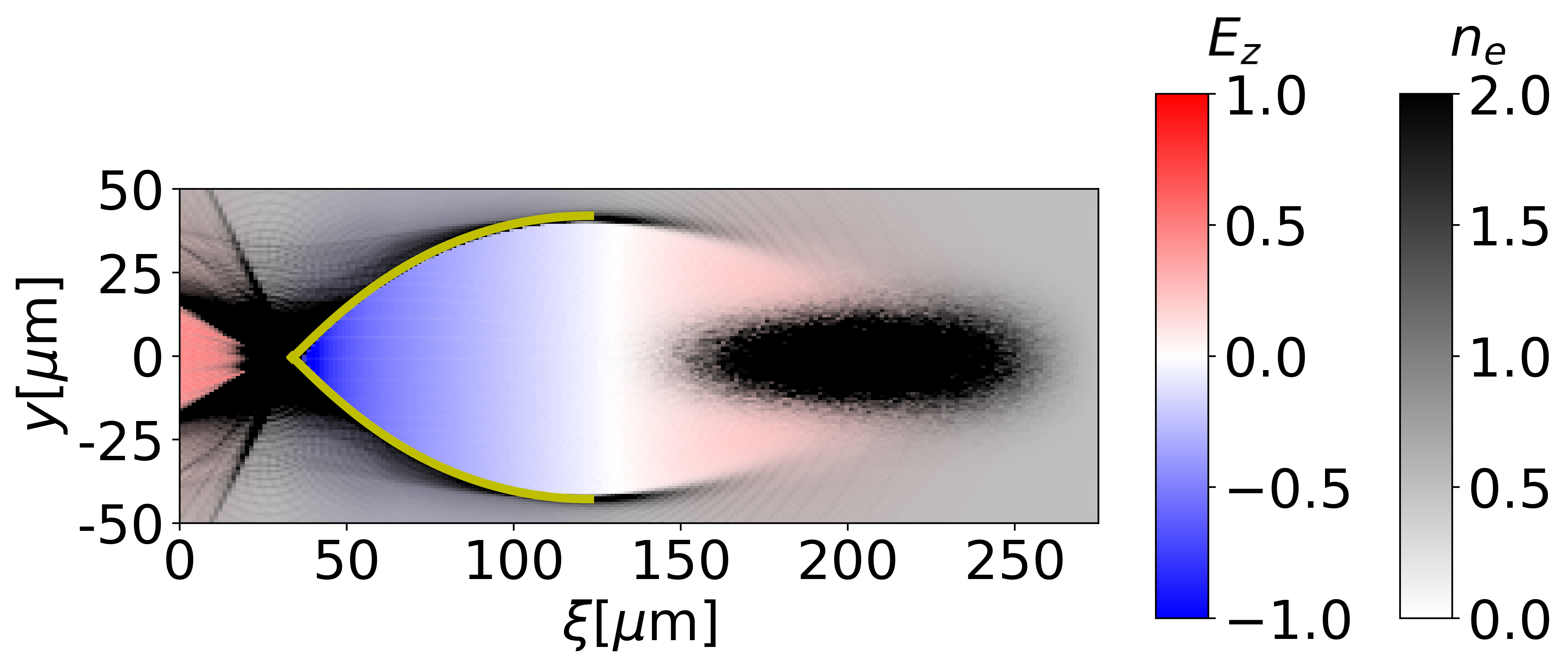}}\\
		\subfloat[]{\label{TroHo_3_layer}\includegraphics[width=\columnwidth]{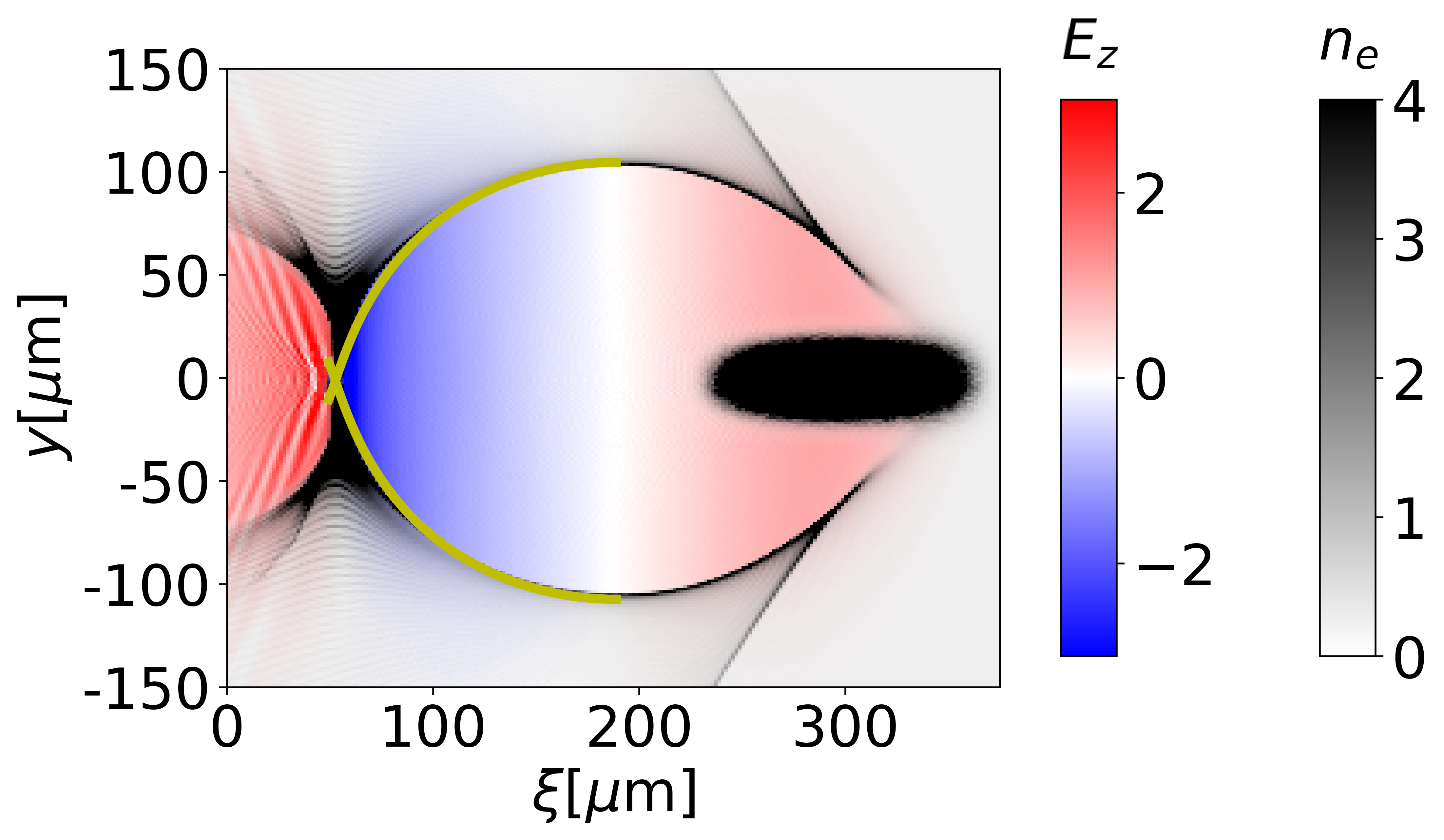}}\\
		\subfloat[]{\label{TroHo_4_layer}\includegraphics[width=\columnwidth]{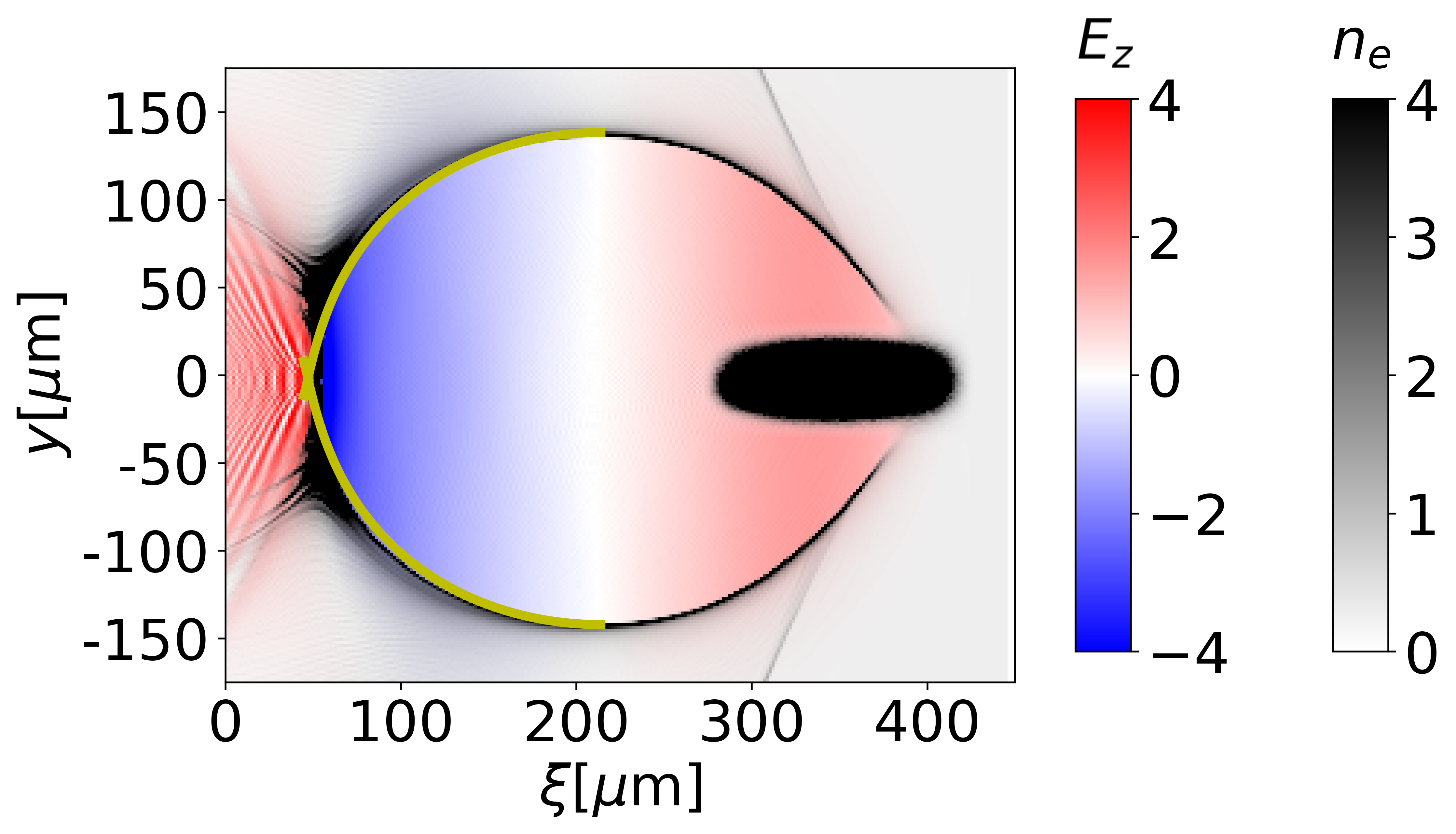}}
		\caption{\label{fig:PICfit}PIC simulation of a bubble in homogeneous plasma, driven by a 1\;GeV bi-Gaussian electron driver with peak density $|\rho_e|=3$ in (\ref{TroHo_0_layer}), $|\rho_e|=24$ in (\ref{TroHo_3_layer}) and $|\rho_e|=48$ in (\ref{TroHo_4_layer}). Shown are electric field in $\xi$-direction and electron density $n_e$. The overlay in yellow shows the good agreement of the analytical model for the electron sheath approximation with the simulations. The plasma wavelength for the simulations is $\lambda_p = $ 200 $\mu$m.}
	\end{figure}
	
	As we can see in Fig. \ref{fig:PICfit}, the correct choice of $\Delta$ and $R$ leads to a perfect fit of the bubble border. The electric field in the separatrix however is not described well using this ansatz (compare Fig. \ref{fig:fieldsBroken}). This is due to the fact that $E_z$ is proportional to $r_b'$, but $r_b'$ is very flat. The strong field in the separatrix leads to these problems of description. This is why we introduce a new methodology in order to fix this effect.

	\subsection{Field modulation near bubble back for homogeneous plasma}
	If we stay within the thin sheath approximation, it is sufficient to use the simplified expression 
	\begin{align}
	& E_z = -S_{I,b}\frac{r_b'}{r_b} \; ,
	\end{align}
	for the longitudinal electric field inside the bubble. If we approximate $E_z$ by its lowest order for small $\xi$ in homogeneous plasmas we see that
	\begin{align}
	& E_z \approx E_0\xi \; , \label{Ez_ana1}
	\end{align}
	where 
	\begin{align}
	& E_0 = \frac{1}{2 + 8 R^{-2}} \; .  
	\end{align}
	In this form $|dE_z/d\xi|$ can't exceed $1/2$, which is the upper limit for very large ($R\gg1$) bubbles. In the same way, $|dE_z/d\xi|$ is limited by zero, which is the lower limit for very small amplitude ($R\ll1$) bubbles.

	If we compare $E_z$ from Eq.~(\ref{Ez_qsm}) to the actual form from PIC simulations in Fig. \ref{fig:fieldsBroken} for homogeneous plasma we see that Eq.~(\ref{Ez_qsm}) does not describe the progression correctly in the rear bubble back. However, the model is in quite good accordance near the bubble mid. The most likely reason for this deviation is the fact that the predicted bubble radius $r_b$ does not decrease strongly enough and that at the bubble rear part so far ignored electron currents become important for the field configuration.
	
	To incorporate the behavior of the $E_z$-field near the bubble stern, we assume that the field diverges at $L=\xi(r_b=0)$. The magnitude of the divergence cannot be reproduced from the current bubble model so we search for a function $f(\xi)$ that bears the divergence but does not change $E_z$ near the bubble mid. A good choice for such a function is
	\begin{align}
	& f(\xi) = \left(\frac{L}{L-\xi}\right)^{1/m} = \frac{1}{\sqrt[m]{1-\frac{\xi}{L}}}
	\end{align}
	which has the desired features 
	\begin{align}
	& \lim_{\xi\rightarrow L}f(\xi) = \infty \; , && f(\xi\approx0) \approx 1 \; .
	\end{align}
	Since $E_z$ from (\ref{Ez_ana1}) is acceptable in a whole range near the bubble mid, we should choose a parameter $m>1$ to reduce the influence of $f(\xi)$ on $E_z$ for too small $\xi$. A rather rough fit to PIC simulations shown in Fig. \ref{fig:fieldsRepaired} yields $m \in [3,4]$ being a good choice. Finally, the quasi-analytical longitudinal electrical field in a stretched bubble is
	\begin{align}
	& E_z = \frac{E_0\xi}{\sqrt[4]{1-\frac{\xi}{L}}}\; . \label{Ez_anaEXP}
	\end{align}
	If we follow the argumentation of section \ref{homogeneous}, we get
	
	\begin{align}
	 B_\varphi &= -\frac{r}{2} \cdot \frac{E_0 \left(4 L - 3 \xi \right)}{4L \sqrt[4]{1-\frac{\xi}{L}}^5} \; ,\\
	 E_r &= \frac{r}{2}\left(1-\frac{E_0 \left(4 L - 3 \xi \right)}{4L \sqrt[4]{1-\frac{\xi}{L}}^5} \right) \; , \\
	 F_z &\approx -\frac{E_0\xi}{\sqrt[4]{1-\frac{\xi}{L}}} \; ,\\
	 F_r &= -\frac{r}{2} +\left(1 -\frac{p_z}{\gamma}\right)\frac{E_0 \left(4 L - 3 \xi \right)}{4L \sqrt[4]{1-\frac{\xi}{L}}^5} \; . \label{Er_ana1}
	\end{align}

	
	\section{Conclusion}
	Current analytical models of the bubble or blow-out regime of plasma wakefield---while being very successful at describing the boundary of the bubble---fail to describe the divergence of the accelerating electric field at the wakefield rear in the bubble for extremely small amplitude wakefields which are relevant for Trojan horse injection regime of PWFA.
	In this paper, we have derived a phenomenological theory that fixes this problem using a repair function.
	The repair function was chosen to correspond to the results of PIC simulations for bubbles of different size in the case of homogeneous plasmas.
	This model can be used to study the acceleration of electrons in physically correct accelerating and focusing fields.
	However, the model does not consider the effect of beam loading on the distribution of the electric fields. 
	Also, while available analytical models are applicable to plasmas with arbitrary radial density distributions, it is unclear whether the proposed repair function will hold cases of inhomogeneous radial profiles.
	Further research is required to investigate these problems.
	\begin{acknowledgments}
		This work has been supported in parts by DFG (project PU 213/6-1) by BMBF (project 05K16PFB), by the Russian Science Foundation (project No. 18-11-00210, A.G. and I.Yu.K.), and the Russian Foundation for Basic Research (projects No. 20-52-50013 and 20-02-00691).
	\end{acknowledgments}
	
	\bibliographystyle{unsrt}

\end{document}